\documentclass[twocolumn,aps,showpacs,preprintnumbers,amsmath,amssymb]{revtex4}

\usepackage[pdftex]{graphics}
\usepackage{dcolumn}
\usepackage{bm}
\usepackage[latin1]{inputenc}

\begin{document}

\newcommand{\SiO}{SiO$ _2$}
\newcommand{\LM}{optical microscope}


\title{Morphology and flexibility of graphene and few-layer graphene on various substrates}

\author{U. Stöberl}
 \email{Ulrich.Stoeberl@physik.uni-r.de}
 \affiliation{Institut für Experimentelle und Angewandte Physik, Universität Regensburg, 93040 Regensburg, Germany}

\author{U. Wurstbauer}
 \altaffiliation{present address: Department of Physics, University of Hamburg}
 \affiliation{Institut für Experimentelle und Angewandte Physik, Universität Regensburg, 93040 Regensburg, Germany}

\author{W. Wegscheider}
 \affiliation{Institut für Experimentelle und Angewandte Physik, Universität Regensburg, 93040 Regensburg, Germany}

\author{D. Weiss}
 \affiliation{Institut für Experimentelle und Angewandte Physik, Universität Regensburg, 93040 Regensburg, Germany}

\author{J. Eroms}
 \affiliation{Institut für Experimentelle und Angewandte Physik, Universität Regensburg, 93040 Regensburg, Germany}

\date{\today}

\begin{abstract}
\noindent
We report on detailed microscopy studies of graphene and few-layer-graphene produced by mechanical exfoliation on various semi-conducting substrates. We demonstrate the possibility to prepare and analyze graphene on (001)-GaAs, manganese p-doped (001)-GaAs and InGaAs substrates. The morphology of graphene on these substrates was investigated by scanning electron and atomic force microscopy and compared to layers on \SiO. It was found that graphene sheets strongly follow the texture of the sustaining substrates independent on doping, polarity or roughness. Furthermore resist residues exist on top of graphene after a lithographic step. The obtained results provide the opportunity to research the graphene-substrate interactions.
\end{abstract}

\maketitle
\noindent
Since the discovery of graphene sheets in 2004 a wealth of unusual properties of this gapless semiconductor has been explored experimentally \cite{Novoselov_1st-Science,Novoselov_PNAS-2D-Crystals,Novoselov_2d-massless-DiracFermions,Zhang-Kim_nature,Berger_JPhysChemB_EpitaxialGraphene}.
Theoretically, graphene as a building block of graphite has been studied extensively, starting from the 
middle of the past century \cite{Wallace_PR_GraphitBandTheory,McClure_PR_GraphiteDiamagnetism,Slonczewski_PR_GraphiteBandStructure}. Even though 3D graphite is an abundant material with many applications, the way to 2D graphene took a long time \cite{Berger_JPhysChemB_EpitaxialGraphene,Geim_naturMat_RiseOfGraphene}. The start to this area of research was the discovery that mechanically exfoliated graphene sheets are visible under an optical microscope, if an oxidized silicon wafer with a certain thickness of oxide is used as a substrate. Nowadays, even the number of graphene layers can be determined by optical inspection. Therefore, most experimental studies rely on oxidized silicon as a substrate. Spatially resolved Raman spectroscopy can distinguish between single layer, bilayer and multilayer graphene and was also performed to study the influence of the substrate on the Raman scattering spectrum \cite{APL-Raman-ConGlass,JoPhy-Raman-ConGlass} of graphene by investigating the graphene-substrate phonon coupling. The substrate was shown to be limiting the carrier mobility in experiments with freely suspended graphene sheets \cite{Kim_PRL_SuspendedGraphene,Du_conMat_SuspendedGraphene}, where the mobility was increased by a factor of ten. While this clearly works as an impressive proof of concept, device applications requiring high mobilities cannot be realized in this way, and alternative substrates need to be explored. For instance, GaAs or InGaAs substrates, where the dielectric constant is much higher than in \SiO{}, could have shorter screening lengths of charged impurities. Further possibilities of taking advantage of a suitable choice of substrate were suggested in recent theory articles. For example, single layer graphene deposited on a boron nitride surface could develop an energy gap \cite{Giovannetti_PRB_GapThroughBorNitride}. Using a manganese doped substrate for bilayer graphene preparation should also lead to a gap and even to a highly spin polarized state \cite{Mao_Nanotechnology_MnWithBilayer}, which would be ground-breaking for spintronics in carbon-based devices.

\noindent
In this letter we report our experimental studies on the influence of different kinds of substrates on the morphology of graphene and few layer graphene using scanning electron microscopy (SEM) and atomic force microscopy (AFM). We investigate the polar (001) semi-insulating GaAs surface (I), Mn-doped (001) GaAs (II), In$_{0.75}$Ga$_{0.25}$As (III), and 300nm \SiO~(IV) as a reference. In this way we are able to change the polarity, the local electromagnetic field and the texture of the substrate surface. Apart from the oxidized Si wafer, all substrates are grown by molecular beam epitaxy on standard n-doped or undoped (001) GaAs substrates. Substrate (I) consists of a semi insulating GaAs layer, substrate (II) of a manganese p-doped GaAs layer with a carrier concentration in the range of $5 \times 10^{19} \mathrm{cm}^{-3}$, and substrate (III) is an In$_{0.75}$Ga$_{0.25}$As layer grown on a metamorphic buffer layer for strain relaxation, which leads to a strongly corrugated (cross-hatched) surface morphology with a height variation up to 14 nm and a periodicity of about 1$\mu$m. The MBE growth of the substrates is described elsewhere \cite{Wurstbauer_JofCrystalGrowth_GaMnAs,Wurstbauer_PhysicaE_InGaAs}. The reference substrate (IV) is a 300nm thick \SiO~layer on highly doped silicon.\\
The graphene and few-layer graphene (FLG) sheets were deposited using the micromechanical cleavage technique with natural bulk graphite as it has been introduced in \cite{Novoselov_1st-Science,Novoselov_PNAS-2D-Crystals,Zhang-Kim_nature}. We want to point out that we did not perform lithography on the substrates prior to  graphene deposition in order to avoid changes of the surface properties and to exclude unwanted residues underneath the graphene sheets. For the \SiO~substrate (IV) the identification of the graphene and FLG sheets was done by an \LM. On all other substrates the sheets were found by searching the entire chip surface with a SEM operating at a few kV and equipped with an in-lens detector. 
After identification of suitable flakes poly methyl methacrylate (PMMA) resist is spun onto the samples and one electron-beam lithography (EBL) step is performed to define alignment marks close to the flakes. This facilitates locating the flakes in the AFM. 
After metal evaporation lift-off is performed by a warm acetone bath followed by an isopropanol rinse. The EBL step was also performed on reference samples of all substrates without graphene. The AFM measurements were done under ambient conditions with a MultiMode IIIa AFM in intermittent contact-mode with standard silicon tips.

\begin{figure}
\centering
\includegraphics{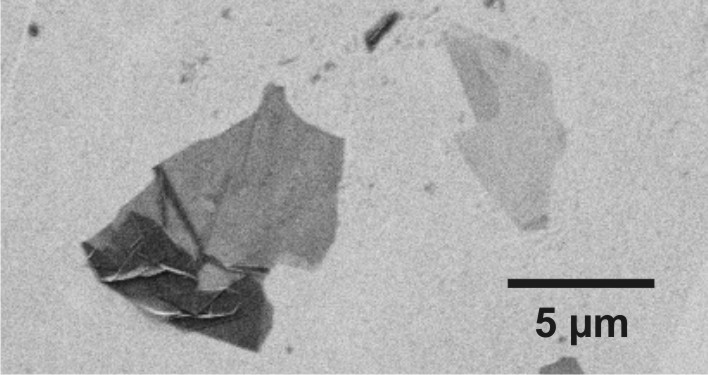}
\caption{SEM image of a thin-graphite and a few-layer-graphene flake on substrate (I). The flake on the right side is 4~nm high, the left one 15~nm.}
\label{fig:GaAs-REM}
\end{figure}

\noindent
Figure \ref{fig:GaAs-REM} shows an example of a typical SEM picture during the identification process. Two flakes on substrate (I) can be clearly distinguished. On the right flake one can also detect the folding of the few-layer-graphene (FLG) in the upper left corner. The right flake has an AFM step height of $4$~nm, and of $8.7$~nm in the folded region. The small difference of $0.7$~nm is presumably due to imperfect stacking of the folded region and the fact that more than one sheet is folded. The left flake has a height of about 15~nm and appears much darker in the SEM image. The SEM contrast allows us to distinguish graphene/few-layer graphene from thicker, graphitic flakes on all substrates, while the precise number of layers has to be determined by other means.

\begin{figure}
\centering
\includegraphics{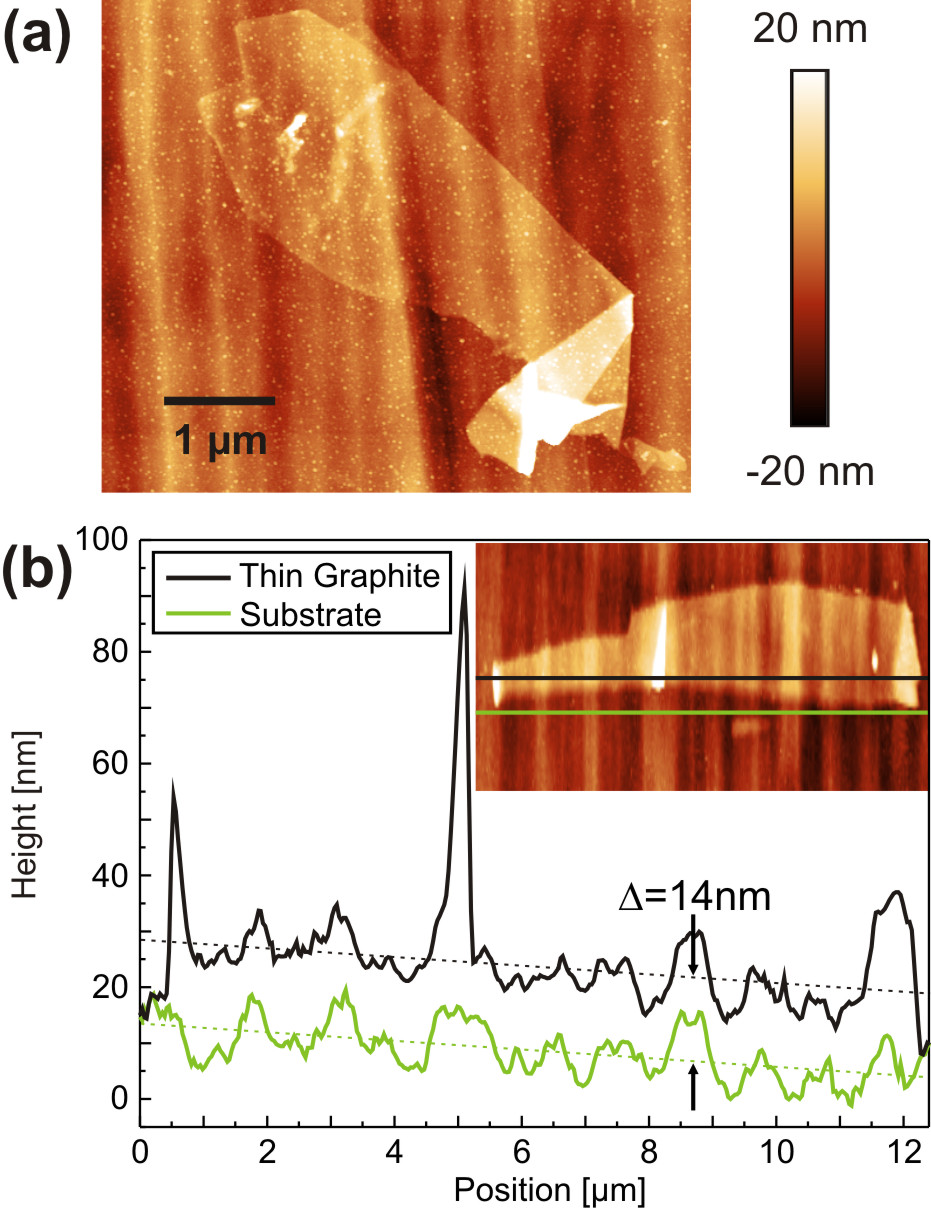}
\caption{(a): AFM image of graphene on substrate (III). The height is $1.6$~nm. (b): Line-cuts through substrate and a $14$~nm thin graphite flake. The insert shows this flake. The y-axis is stretched by a factor 3. The peak at  $\sim 5 \mu$m is due to some layers pointing upwards.}
\label{fig:InGaAs-AFM}
\end{figure}

\noindent
The AFM investigations of the flakes give us additional information. In Figure \ref{fig:InGaAs-AFM} (a), a $1.6$~nm thick flake on substrate (III) is shown. Since the AFM step height of a graphene single layer on \SiO~was reported to vary between $\sim 5$~\AA~and $\sim 10$~\AA~ \cite{Moser-EFM-APL,Graf_nanoLett_SpatialResRaman}, and the layer separation in graphite is $3.35$~\AA, we estimate the number of layers of the flake between 2 and 4. Substrate (III) has a strong corrugation, resulting from the step-graded buffer in the MBE growth, and the AFM image clearly shows that the flake follows the substrate texture closely. Thicker flakes also demonstrate this pronounced flexibility. Figure~\ref{fig:InGaAs-AFM} (b) shows data taken on a $14$~nm thick graphite flake, again on substrate (III). The line scan through the flake shows a quite accurate and detailed copy of the underlying substrate texture even for such a large number of graphene layers. We observed this behavior on all samples and all substrates with a graphite thickness of up to 40~nm. In the following, we concentrate on flakes with a thickness $d \leq 2$~nm.

\begin{figure}
\centering
\includegraphics{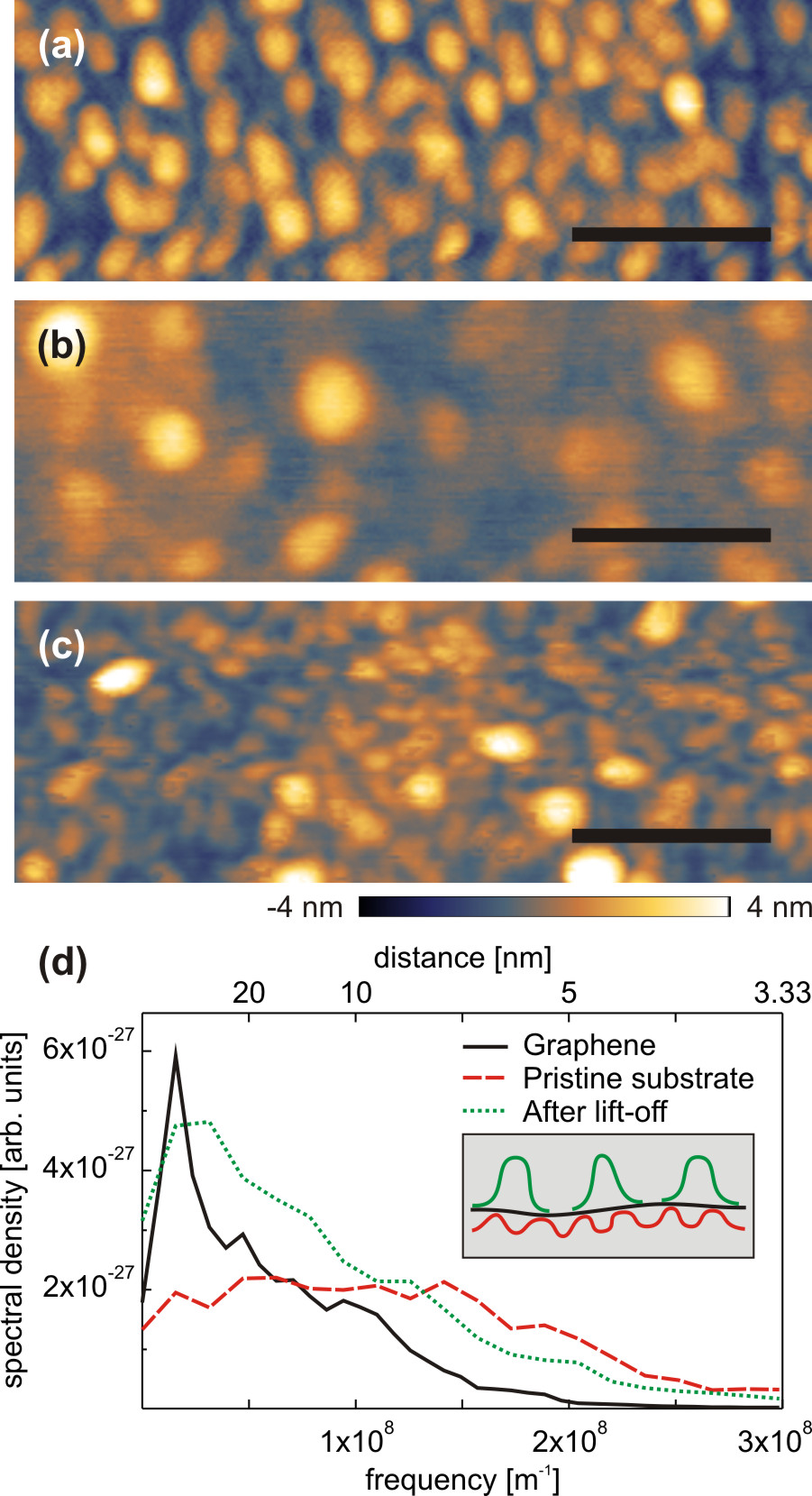}
\caption{AFM-images of (a) pristine substrate (I) (undoped GaAs), (b) graphene on substrate (I), (c) substrate (I) after liftoff. (d): power spectral density of AFM-images, the insert schematically drafts the situation. All scale bars 100~nm}
\label{fig:GaAs-AFM-Analysis}
\end{figure}

\noindent
To gain more insight into the flexibility of graphene sheets and the influence of the morphology of the substrates, we performed detailed AFM scans of graphene flakes and of the corresponding substrates. To check for unwanted process residues on the substrates, we also compared the pristine substrates and the substrates after e-beam lithography. Since the mechanical exfoliation technique with adhesive tape always leaves residues on the surface \cite{Moser-EFM-APL}, we could not use the areas outside the graphene flakes as a reference. Instead, we prepared control samples with the same lithographic processing, but without depositing graphene flakes. Typical AFM images are reproduced in Fig.~\ref{fig:GaAs-AFM-Analysis}. In (a) a close up AFM scan of the pristine substrate is given. The morphology of a graphene sheet ($d \approx 1.3$~nm) on substrate (I) is displayed in (b). In (c) a control sample without graphene after lift-off processing is shown. The roughness of the pristine substrate shown in Fig. \ref{fig:GaAs-AFM-Analysis} (a) does not appear in the image of the graphene flake (b). The flake seems to cover this fine texture originating from the substrate. But the inspection of the graphene surface displayed in (b) reveals noticeable singular spots, which are about $4$~nm high and $30$~nm in diameter. Similar spots appear on the reference samples after lift-off processing, but not on the pristine samples. Therefore, we conclude that those spots in (b) must be PMMA residues, which lie on top of the graphene flakes, since the alignment markers are deposited after graphene preparation. A sketch of the situation is inserted in Fig. \ref{fig:GaAs-AFM-Analysis} (d). Note that those spots appear even though the PMMA was carefully removed in the lift-off procedure with warm acetone. We cannot use strong oxidizing agents or oxygen plasma for more thorough cleaning since this could also damage the graphene flakes. Similar results (on \SiO-substrates) were reported in \cite{Ishigami_nanoLett_STM-GonSiO2}.

\noindent
We were also interested in the spatial frequency dependence of the flexibility of the graphene sheets. Since the substrates have an intrinsic surface roughness extending to high spatial frequencies, we use this as a ``test signal'' to probe how closely the graphene flakes follow the underlying structure. Therefore we calculated the power spectral density of the AFM images of graphene flakes and of the underlying substrate, before and after lift-off processing. Each curve is the rms average of the 1D power spectral densities of the individual line scans of an AFM image. For the pristine and the resist covered one the curves of several $400\times 200$~nm$^2$ areas were averaged to reduce the scatter in the data. The graphene curve was calculated from an $800 \times 800$~nm$^2$ image. The pristine GaAs surface has a rather flat spectrum up to $1.6\times 10^8$~m$^{-1}$ which then gradually rolls off at higher spatial frequencies, presumably due to the limited lateral resolution of the AFM. The graphene surface shows a similar spectral density at intermediate frequencies, but the roll-off occurs already at $1.2\times 10^8$~m$^{-1}$ (corresponding to $\sim 8$~nm) and is somewhat steeper. This defines the cut-off frequency up to which the graphene follows the substrate corrugations. For higher frequencies the stiffness of graphene prevents the sheet from following the substrate in total \cite{Ishigami_nanoLett_STM-GonSiO2}. The pronounced peak at low frequencies is found on the graphene samples and on the substrates after lift-off processing, but not on the untreated substrate, and hence is again due to the PMMA dots on top of the graphene. From these data and the fact that we also found these spots on graphene on \SiO~after an EBL step but not if there is not such a step we conclude that there are some residues on the surface of graphene after lift-off.

\noindent
To conclude we show that graphene and few layer graphene can routinely be deposited and detected on different kinds of substrates, namely (001)-GaAs, Mn p-doped GaAs and InGaAs and compared this to \SiO~as a well known reference. On all of these substrates, graphene with $d \leq 1.6$~nm has been detected and investigated. Therefore the various electromagnetic substrate configurations of our substrates ranging from amorphous \SiO~over polar (001) GaAs to p-doped GaMnAs do not affect the formation or stability of graphene. We also demonstrate that graphene follows a continuous substrate texture from $\sim 8$~nm on up to more than $1 ~\mu$m. Finally, the AFM images and power spectral densities show that even with careful lift-off, an EBL step leaves unwanted PMMA residues on the surface.
The effect of the substrate on the electronic properties of graphene and the comparison of the step-height of one monolayer on these materials is part of ongoing investigations.\\
This work was supported by the DFG via GK 638.

\end{document}